\documentclass{article}

\usepackage[preprint]{neurips_2024}

\usepackage[utf8]{inputenc} 
\usepackage[T1]{fontenc}    
\usepackage{hyperref}       
\usepackage{url}            
\usepackage{booktabs}       
\usepackage{amsfonts}       
\usepackage{nicefrac}       
\usepackage{microtype}      
\usepackage{xcolor}         

\usepackage{graphicx}
\usepackage{algorithm}
\usepackage{algorithmic}
\usepackage{amsthm,amssymb,amsmath}
\usepackage{makecell}
\usepackage{multirow}

\usepackage{caption}
\usepackage{subcaption}

\title{Why Perturbing Symbolic Music is Necessary: Fitting the Distribution of Never-used Notes through a Joint Probabilistic Diffusion Model}

\author{
Shipei Liu\\
School of Software, \\
Dalian University of Technology\\
Dalian 116621, China\\
\texttt{tapliu@mail.dlut.edu.cn} \\
\And
Xiaoya Fan \\
School of Software, \\
Dalian University of Technology\\
Dalian 116621, China\\
\texttt{xiaoyafan@dlut.edu.cn} \\
\AND
Guowei Wu \\
School of Software, \\
Dalian University of Technology\\
Dalian 116621, China\\
\texttt{wgwdut@dlut.edu.cn} \\
}

\begin{document}

\maketitle

\begin{abstract}
Existing music generation models are mostly language-based, neglecting the frequency continuity property of notes, resulting in inadequate fitting of rare or never-used notes and thus reducing the diversity of generated samples.
We argue that the distribution of notes can be modeled by translational invariance and periodicity, especially using diffusion models to generalize notes by injecting frequency-domain Gaussian noise.
However, due to the low-density nature of music symbols, estimating the distribution of notes latent in the high-density solution space poses significant challenges.
To address this problem, we introduce the Music-Diff architecture, which fits a joint distribution of notes and accompanying semantic information to generate symbolic music conditionally.
We first enhance the fragmentation module for extracting semantics by using event-based notations and the structural similarity index, thereby preventing boundary blurring.
As a prerequisite for multivariate perturbation, we introduce a joint pre-training method to construct the progressions between notes and musical semantics while avoiding direct modeling of low-density notes.
Finally, we recover the perturbed notes by a multi-branch denoiser that fits multiple noise objectives via Pareto optimization.
Our experiments suggest that in contrast to language models, joint probability diffusion models perturbing at both note and semantic levels can provide more sample diversity and compositional regularity.
The case study highlights the rhythmic advantages of our model over language- and DDPMs-based models by analyzing the hierarchical structure expressed in the self-similarity metrics.
\end{abstract}

\section{Introduction}
With the emergence of music generation applications like SunoAI, artificial intelligence technology is thriving in music composition.
Typically, symbolic music generation is perceived as a sequence generation task, such as VAE-based models~\cite{TransformerVAE}, which heavily rely on surrogate objectives to approximate maximum likelihood during training;
and Transformer-based models~\cite{Transformer-GANs, HuangY20}, which encounter challenges such as error accumulation in long sequences and the issue of quadratic growth in time complexity.
However, it has come to our attention that these language-based music generation models overlook the continuous property of notes. 
For instance, the latest RWKV-music~\cite{PengAAAABCCCDDG23} initializes all 128 notes but does not actively update the distribution of notes outside the training corpus.
Therefore, we argue that modeling musical symbols differs from language, as distributions of "never-used" notes can be deduced through frequency characteristics, such as translational invariance.

Diffusion models excel at intuitively learning data distributions, so we can leverage them to fit distributions of "never-used" notes that have been obtained by generalizing in the frequency domain.
Existing research on music diffusion models falls into two categories.
On one hand, researchers directly applied continuous diffusion models to acoustic spectrograms.
Typical works like DiffRoll~\cite{CheukSUMTTHM23} and MAID~\cite{LiuGY23a} take either spectrograms or paired MIDI-spectrograms as input, generating music phrases by recovering noise-added audio.
These models' performance is unsatisfactory, facing the challenge of misalignment between frequency-domain audio and notes, known as the "out-of-tune" problem.
On the other hand, researchers mapping note symbols as continuous latent variables, thus applying discrete diffusion models.
For example,~\cite{MittalEHS21} utilized the diffusion model after parameterizing notes.
However, this approach heavily relies on the performance of MusicVAE~\cite{RobertsERHE18} and post hoc classifier guidance, leading to an unstable generation process.
~\cite{PlasserPW23} addresses these issues by employing Denoising Diffusion Probabilistic Models (DDPMs) to recover corrupted notes and thus generate music.
Unfortunately, these discrete-continuous mapping diffusion models overlook the low-density nature of the data, resulting in lower diversity in generated samples.

Although diffusion models can be applied to music generation through appropriate methods, they face the challenge of striking a trade-off between diversity and regularity.
To overcome the constraint problem induced by classifier guidance,~\cite{Zhu0ORT023} proposed a contrastive diffusion model, exploring the relationship between diffusion training and diverse objectives in dance-to-music tasks.
~\cite{li2024music} introduces time-varying text inversion and bias-reduced stylization techniques to capture different levels of melodic-spectrogram features, during music style transfer.
Additionally, some methods gain ground in generating structured waveforms or spectrogram music conditioned on text prompts, such as ERNIE-Music~\cite{abs-2302-04456} and Noise2Music~\cite{abs-2302-03917}.
These works suggest that diffusion models can reflect music semantics in samples through conditioning or prompts, thereby limiting the irregularity of generated samples.

We consider two typical problems of diffusion-based music generation models: (1) maintaining continuity when embedding low-density notes into high-density surrogates; and (2) conditional denoising process with semantic prompt.
In this paper, we introduce a Music-Diff architecture that utilizes hierarchical music semantics for multivariate diffusion, followed by sequentially generating music phrases with conditional prompts.
Our novelty lies in perturbing and denoising joint distribution, thereby generalizing finite discrete notes in the solution space, and alleviating the difficulty of estimating distributions in low-density regions.
To summarize, our contributions include:

\begin{itemize}
\item Going beyond previous methods, we improve the fragmentation module by basing the candidate window on event-based notations, resulting in more precise segmentation of structural music elements.
Moreover, we assign greater emphasis on fuzzy boundaries through the structural similarity index, thereby significantly reducing fine-grained deviations.
\item We propose a non-language tokenizer and a joint semantic pre-training method, followed by multivariate perturbation of the joint probabilities of notes, chords, and sections.
When extending notes to continuous surrogates, these methods play a crucial role in maintaining the continuous musical properties of high-density spaces.
\item We present a parallel denoiser for recovering perturbed notes with semantic prompts, which employs a Pareto optimization.
Compared to language- and DDPMs-based models, the generated samples exhibit greater diversity and structural regularity within our experiments.
\end{itemize}


\section{Related Work}
In this section, we review the literature on language- and diffusion-based music generation models.

\textbf{Language-based models:}
Many works on music modeling employ structural segmentation techniques, e.g., using hierarchical neural networks for unsupervised music segmentation to detect boundaries of music segments~\cite{Segmentation, WangSW19}.
Subsequently, language-based music generation models, such as Muse-BERT~\cite{wang2021musebert}, aim to encode music through its structure and polyphonic characteristics, thus leading to melody reconstruction compared to other models.
To standardize structural segmentation results,~\cite{de2022measuring} conducted hierarchical segmentation experiments on music and proposed quantitative metrics for computing structural components.
Most effectively, a Transformer-based model~\cite{li2023fine} captures the interaction between musical elements and generates event-based sequences through a music event position encoding scheme.
Language-based models have also made progress in improvisational music composition, e.g., a Transformer-based MMT model~\cite{dong2023multitrack}, which learns multi-track music collaboration, has achieved commendable results in subjective listening tests.
Furthermore, some scholars have considered the importance of continuous attributes such as translational invariance in music. 
For instance, the FME model~\cite{guo2023domain} embeds music space through relative pitch, interval, and onset embeddings, thereby leveraging music knowledge to model symbolic music.

\textbf{Diffusion-based models}
Due to the oversight of data continuity in language-based approaches, we strive to explore music diffusion models in this paper.
However, diffusion models are usually applied to continuous data, which is related to their theoretical foundations, non-equilibrium thermodynamics, and score matching~\cite{song2019generative}.
For example, a cascaded latent diffusion approach~\cite{schneider2023mo} directly utilizes waveform data and textual conditions to generate music audio, but this continuous-domain method is prone to pitch-shifting phenomena.
Therefore, more researchers, e.g.~\cite{ji2023emomusictv}, project music symbols to continuous spaces, then perturb and denoise high-dimensional variables, and finally reduce the variables to notes through VAE decoders.
To maintain the music properties in the continuous space, a diffusion model~\cite{mariani2023multi} matches discrete and continuous data in a joint latent space but has the disadvantage of relying heavily on the solution of the variational lower bound variant.
Furthermore,~\cite{HawthorneSRZGME22} compared autoregressive and diffusion models in music generation and found that the diffusion model was superior in both Fréchet distance and reconstruction quality.
    
\section{Proposed method}
\begin{figure}
\centering
\includegraphics[width=0.98\columnwidth]{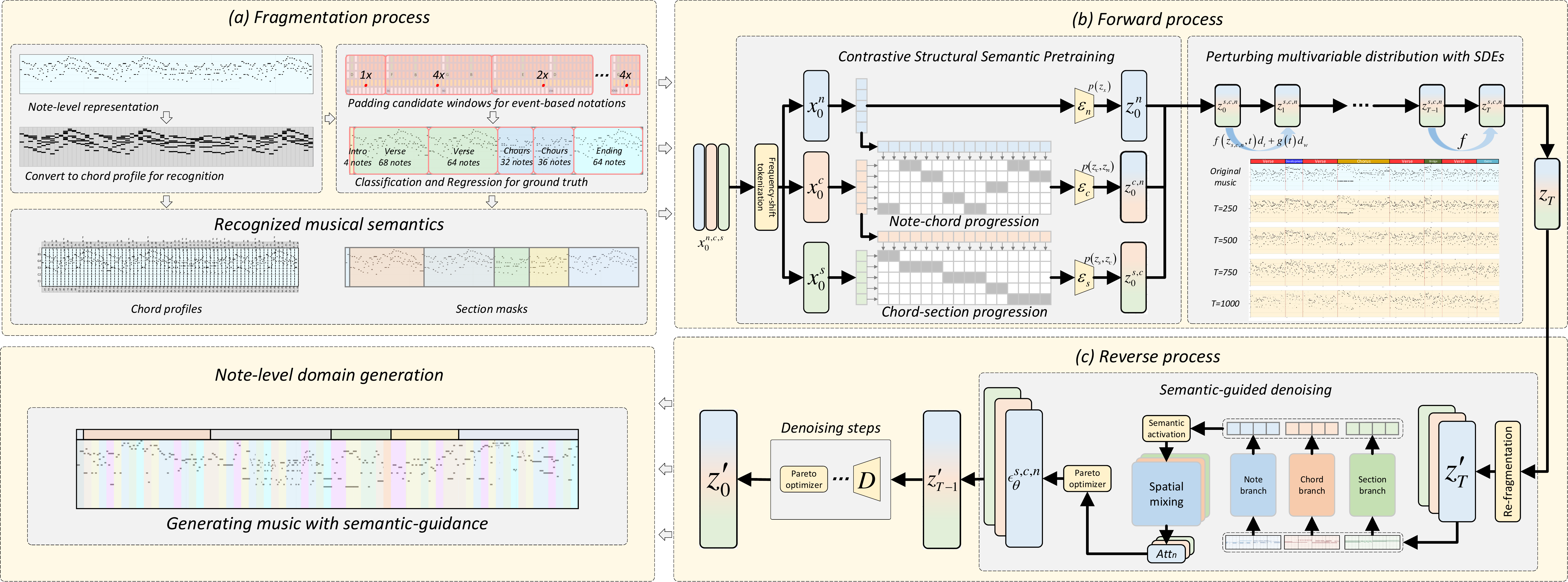}
\caption{Illustration of our Music-Diff architecture, consisting of (a) fragmentation, (b) forward, and (c) reverse processes for semantic extraction, noise perturbation, and note recovery, respectively.}
\label{fig1}
\end{figure}

Our core concept aims to enrich sample diversity through diffusion processes while upholding compositional regularity by harnessing musical structure. 
The proposed Music-Diff framework, depicted in Fig.~\ref{fig1}, comprises:
(a) Fragmentation process, utilize event-based notation and the SSIM index to prevent note-level boundary deviations;
(b) Forward process, introducing joint semantic pre-training to grasp note-semantic progressions and diffuse note distribution via multivariate perturbations;
(c) Reverse process, introducing a multi-branch Symb-RWKV, utilizing Pareto optimization constraints for the conditional generation.
Explanations of Symbols in the paper are detailed in Appendix A.

\subsection{Structural fragmentation challenge: musical semantic matters}
We delve into the weaknesses of current music structure segmentation models and integrated event-based notations and structural similarity index to reduce fine-grained bias.

\textbf{Events-based notations:} 
The minimum unit we used is the MIDI note, which includes 128 dimensions satisfying the twelve-tone equal temperament [\textbf{C}, C$\sharp$, \textbf{D}, D$\sharp$, \textbf{E}, \textbf{F}, F$\sharp$, \textbf{G}, G$\sharp$, \textbf{A}, A$\sharp$, \textbf{B}].
We need to further recognize chords and musical events on the MIDI score to understand the musical higher-order semantics.
We've previously summarized 48 chord profiles~\cite{wu2023power} for recognizing chords by counting the notes within each bar. 
Meanwhile, we extract higher-order semantics through event-based notations, e.g., REMI~\cite{HuangY20} or Compound words~\cite{hsiao2021compound}.
To distinguish musical events, we represent them based on rhythmic or chordal changes, which reduces sequence length while preserving structural information.

\textbf{FSL-V2 Module:} 
The modified fragmentation module searches structural boundaries at the event level instead of the note level, which preserves more structural integrity and narrows the retrieval range.
Specifically, given the event-based notations $x_{cp}$, we set randomized centers $C_{m}$ and fill $m$ non-overlapping windows.
The sizes $\eta_{m}$ of padded windows are selected as a multiple of the bar length (16 semiquavers).
The state transition function of the candidate windows can be written as
\begin{equation}
\{Cand\}^{M}=\mathop{\arg\max}\sum_{m}^{M}p_{cls}(x_{cp}[C_{m}\pm\frac{1}{2}\eta_{m}]),
\label{eq1}
\end{equation}
for all satisfies $(\eta_{m}+\eta_{m-1})/2<C_{m}-C_{m-1}$.
The candidate windows are padded by maximizing the average confidence $p_{cls}$, which are calculated by the classification probabilities.

\textbf{Hybrid Loss:}
We use the structural similarity index (SSIM)~\cite{wang2003multiscale} to eliminate fine-grained bias and thus improve the accuracy of recognizing fuzzy boundaries of musical structures.
Since the SSIM index increases with regional differences, assigning higher weights to fuzzy boundaries can improve segment accuracy.
By clipping the corresponding scope from the predicted candidates and ground truth section, SSIM losses $\mathcal{L}_{SSIM}$ are defined as,
\begin{equation}
\mathcal{L}_{SSIM}(cand,s)=1-\frac{(2\mu_{cand}\mu_{s}+\delta_{1})(2\sigma_{cand,s}+\delta_{2})}{(\mu_{cand}^{2}+\mu_{s}^{2}+\delta_{1})(\sigma_{cand}^{2}+\sigma_{s}^{2}+\delta_{2})},
\label{eq2}
\end{equation}
where $\mu$ and $\sigma$ are the mean and standard deviations of pitches within sections;
$\sigma_{cand,s}$ denotes the covariance between $cand$ and $s$;
The two constants $\delta_{1}= 0.01^{2}$ and $\delta_{2}=0.03^{2}$ are added to avoid the instability that occurs when the denominator is zero.
By combining the classification ($\mathcal{L}_{cls}$) and regression ($\mathcal{L}_{reg}$) losses~\cite{wu2023power}, we develop a hybrid loss $\mathcal{L}_{frag}=\frac{1}{M} \sum_{m}^{M}(\mathcal{L}_{cls}+\mathcal{L}_{reg}+\mathcal{L}_{ssim})$ for section fragmentation, which can capture both note- and event-level boundaries of music structures. 
After passing the FSL-V2 module, event-based notations are recognized to a triplet $[n_{i}, c_{j}, s_{m}]$ of note, chord, and section.

\subsection{Forward process: joint distribution perturbing}
This subsection elaborates on perturbing notes through the distribution of musical structures in a joint feature space.
As a prerequisite, we construct a Joint Semantic Pre-training (JSP) module to represent musical progressions by learning cross-referenced chords and semantics.

\textbf{Tokenization with frequency shifts:}
Conventional music diffusion models typically use a language tokenizer for discrete-to-continuous mapping. 
However, the perturbed high-dimensional feature can only be induced back to the trained tokens, thus unable to increase sample diversity.
For example, when trained on the J.S. Bach dataset, the existing language-based model updates parameters for 68 (initialized 128) notes that have appeared. 
We argue that the non-language tokenizer used by our model should conform to the following properties:
(1) Translational invariance, intervals differ by a major second is identical in the embedding space, e.g., $|D_{4}-C_{4}|$ = $|D_{5}-C_{5}|$;
(2) Periodicity, the same notes in different octaves express the same melodic information, e.g., $|C_{5}-C_{4}|$ = $|C_{6}-C_{5}|$. 
Therefore, we can utilize these properties to diffuse notes to sparse pitches, as well as to normalize the pitch of an instance directly in the frequency domain.

\begin{figure}[htbp]
\centering
\includegraphics[width=0.98\columnwidth]{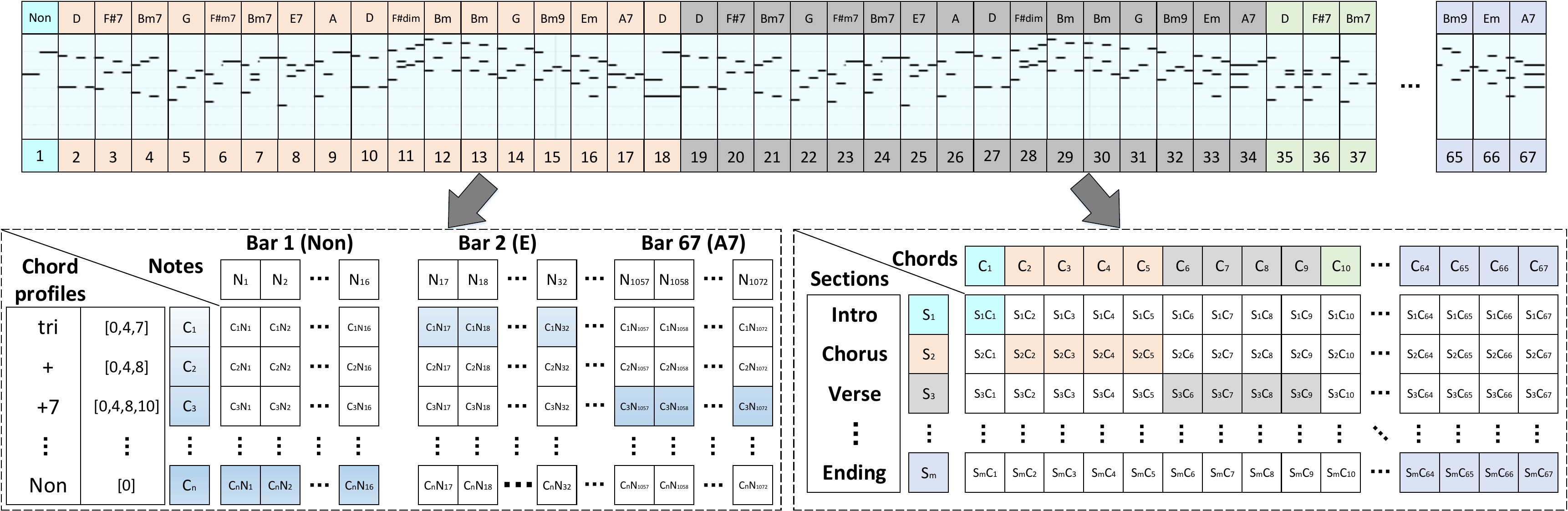}
\caption{Specification of JSP method representing the progression of note-chord and chord-section.}
\label{fig2}
\end{figure}

\textbf{Joint semantic per-training:}
As preparation for multivariate perturbation, we learn the joint distribution by constructing note-chord-section progressions from structural-semantic supervision.
At the same time, we expect this method can expand the low-density tokens to 128 (notes) * 48 (chords) * 10 (sections), similar to language models that typically use word-level tokens (50,277 in RWKV~\cite{PengAAAABCCCDDG23}) instead of just 26 letters.
Assuming that a triplet $[n_{i},c_{j},s_{m}]$, the JSP module predicts which of the possible $i\times j\times m$ pairs will occur, and also expands the variety of embedding vectors.
The note representation $n_{i}$ is embedded in the latent variable $z_{n}$ via an encoder $\varepsilon_{n}(n_{i})$.
By joint progression, we convert the semantics into latent variables $z_{c}$, $z_{s}$ via encoders $\varepsilon_{c}(n_{i},c_{j})$, and $\varepsilon_{s}(c_{j},s_{m})$, respectively.
These latent variables were pre-trained by maximizing the cosine similarity of semantic pairs while minimizing the cosine similarity of randomly incorrect pairs and optimized by symmetric cross-entropy loss.
The JSP procedure is shown in Fig.~\ref{fig2}.

\textbf{Multivariate perturbing:} 
We generalize the idea of perturbing notes with Gaussian noise to multilevel semantics, such that the distribution of perturbed notes evolves according to a deterministic forward process in the joint space.
Thus, we construct a tractable form to generate independent and identically distributed (I.I.D.) samples $z_{0}\langle n,c,s\rangle \rightarrow z_{T}\langle n,c,s\rangle$ indexed by the perturb steps $T$.
To model semantic dependency, we derive joint probabilities between notes, chords, and semantics via Bayes' rule.
Since the conditional distribution can estimate by observations $\mathbb{E}[\epsilon^{c}|c_{t},s_{0}]$ and $\mathbb{E}[\epsilon^{n}|n_{t},c_{0},s_{0}]$, the joint probabilities can be calculate by $p(z_{s,c,n})=\frac{1}{(2\pi)^{n/2}}\exp (-\frac{1}{2}\sum_{i}z_{i}^{2})\propto p(n|c, s)p(c|s)p(s)$.
The diffusion process can be modeled as,
\begin{equation}
d_{z}=\gamma(z_{n,c,s},t)d_{t}+g(t)d_{w},
\label{eq3}
\end{equation}
where $\gamma(\cdot,t)$ and $g(\cdot)$ are the drift and diffusion coefficients obeying a joint Gaussian distribution, respectively.
$d_{w}$ represents the infinitesimal white noise.
By applying the reparameterization trick, we can sample $z_{t} = \sqrt{\alpha_{t}}z_{0} + (1-\alpha_{t})\epsilon_{t}$, where $\epsilon_{t}\sim \mathcal{N}(0, I)$.
The noise perturbation process is formalized by the Markov chain $q(z_{1:T})=\prod_{t=1}^{T}\mathcal{N}(n_{t};\sqrt{\alpha_{t}}(n_{t-1}|s_{0},c_{0}),(1-\alpha_{t})I)$.

\subsection{Reverse process: Joint probabilistic denoising}
The reverse process is to recover the perturbed notes by the probability $p_{\theta}(z_{t})\rightarrow z_{0}$.
Thus, the denoising process can be solved by a \emph{noise prediction network} $f_{\theta}(z_{t},t)$ to learn $q_{\theta}(z_{t-1}|z_{t})$, where $\theta$ is obtained by the reparameterization trick instead of the score-matching method. 

\textbf{Branching denoiser:} 
We construct the perturbation sequence $z_{t}:\langle n_{t},c_{t},s_{t}\rangle$ by re-fragmenting the note sequence for the noise step $t$.
Afterward, we employ a parallel RWKV attention to expedite the denoising process.
Specifically, we utilize multiple semantic-fitting branches to construct a symbiotic denoiser, called Symb-RWKV, aiming at eliminating joint Gaussian noise. 
The introduced denoiser performs WKV~\cite{abs-2105-14103} operations via parallel temporal mixing blocks, and then performs a fusion semantic activation and shared channel mixing blocks.
The attention score is calculated by a weighted sum of the scaled dot-products,
\begin{equation}
Attn(W,K,V)=\frac{\sum_{t=1}^{T-t}e^{W_{t}+K_{T-t}}\odot V_{T-t}}{\sum_{t=1}^{T-t}e^{W_{t}+K_{T-t}}} \in \mathbb{R}^{d},
\label{eq4}
\end{equation}
where $W$ is a trainable parameter representing the position weight decay.
$K$ and $V$ are key, value vectors transformed from parameter matrices.
The output vector posts the WKV operator is implemented using the sigmoid of the receptance $\xi(r)$, computed as $o_{t}=W_{o}\cdot(\xi(r_{t})\odot Attn(wkv_{t}))$.

\begin{figure}[htbp]
\centering
\includegraphics[width=0.98\columnwidth]{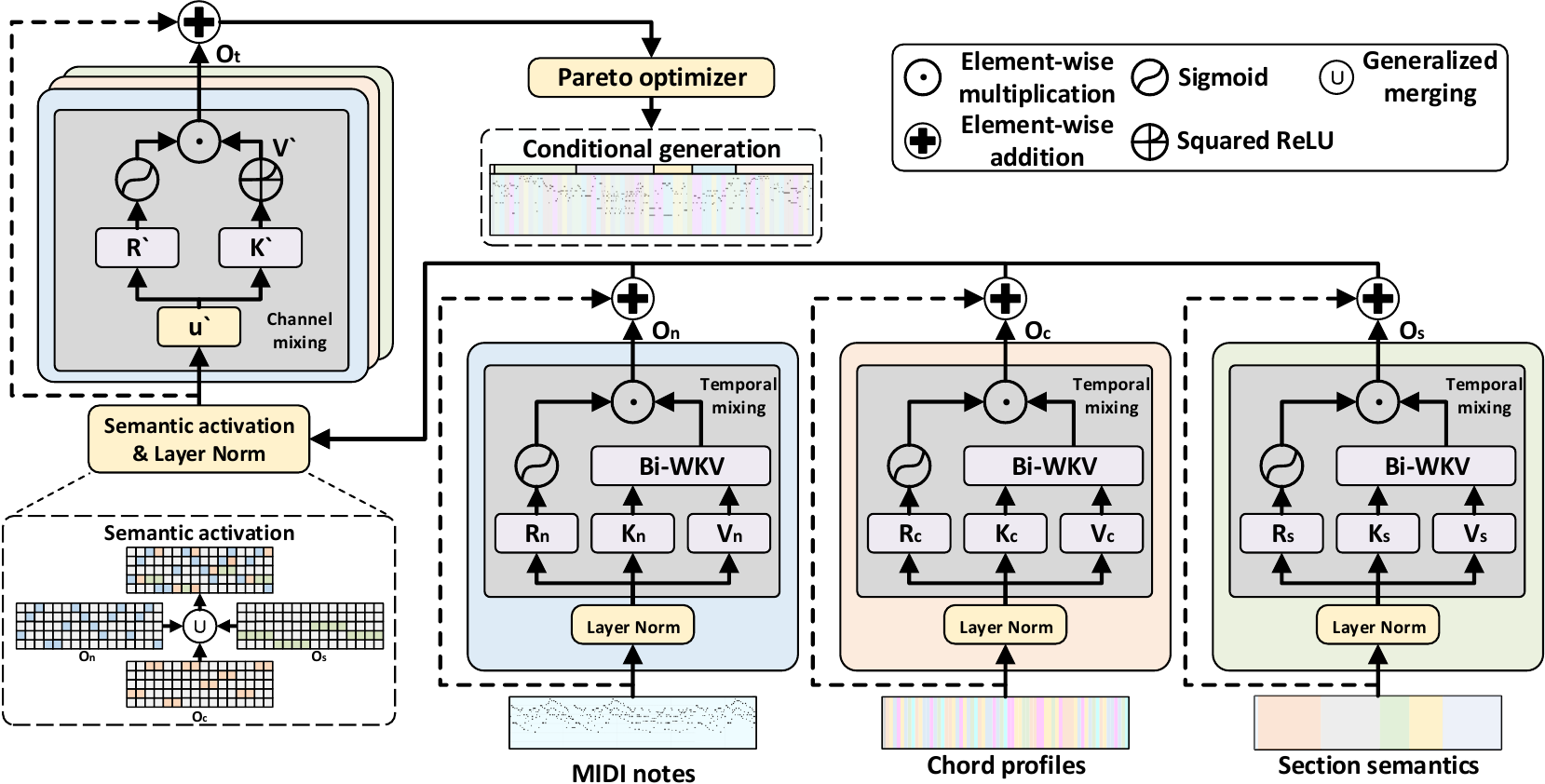}
\caption{Illustration of backbone networks in the denoising.}
\label{fig3}
\end{figure}

To fuse notes and semantics, we design a training-free semantic activation module.
This module generalizes the notes during the training process, denoted as $n_{t}\in \mathbb{R}^{n\wedge s\wedge c}$.
The activation outputs of branches $\varphi\in \phi$ are subsequently subjected to information exchange in the shared channel mix block $o_{t}^{\prime}=\sum_{\varphi}^{\phi}\xi(R_{t})\cdot W_{v,\varphi}\cdot gelu(K_{t,\varphi})\cdot V_{t,\varphi}$, aiming to capture the mutual context.
The prediction layer uses the Gumbel Softmax~\cite{jang2017categorical} $Softmax(o_{t}^{\prime})\to z_{t}^{\prime}$ to predict the noise sequence.
The architecture of Symb-RWKV is shown in Fig.~\ref{fig3}.

\textbf{Semantic prompt:} 
We initialize the model parameters $p(n_{t})\mathcal{N}(s_{0},c_{0}, \boldsymbol{0}; \boldsymbol{I})$ with conditions $c_{0}$, $s_{0}$ and let the note distribution diffuse towards the tractable equilibrium distribution $p(n_{t}|s_{0},c_{0})$. 
Differing from the standard diffusion model, we transfer the objectives from modeling the original distribution $\bigtriangledown_{n_{t}} \log p(n_{t})$ to the conditional distribution $\bigtriangledown_{n_{t}} \log p(n_{t}|c_{0},s_{0})$.
Specifically, we perceive musical patterns through the branches $\epsilon_{\theta}^{c}(c_{t},t)$, and $\epsilon_{\theta}^{s}(s_{t},t)$ by inputting $c_{0}$ and $s_{0}$.
Then, the conditional distribution $\hat{\epsilon}_{\theta}^{n}$ can be fitted by $\epsilon_{\theta}^{n}(n_{t},c_{0},s_{0},t,0)-\epsilon_{\theta}^{n}(n_{t},c_{0},s_{0},t,T)$.
Our model recovers notes from noise by minimizing regression losses,
\begin{equation}
\mathcal{L}_{\theta}=\mathbb{E}_{\epsilon^{n},\epsilon^{c},\epsilon^{s},t}||\hat{\epsilon}_{\theta}(n_{t},s_{0},c_{0},t)-\langle\epsilon^{n},\epsilon^{c},\epsilon^{s}\rangle||^{2}_{2},
\label{eq5}
\end{equation}
where $\langle\cdot\rangle$ denotes concatenation;
The updated semantic $\epsilon_{\theta}^{c}$ and $\epsilon_{\theta}^{s}$ was obtained through re-fragmentation operations.
For simplicity, we denoted the concatenation of $\langle\epsilon_{\theta}^{n},\epsilon_{\theta}^{c},\epsilon_{\theta}^{s}\rangle$ as an output $\hat{\epsilon}_{\theta}$.
Our method attempts to fit all semantic distributions by a joint noise prediction network, requiring that the backbone can handle the mutual interaction between semantic conditions.

\textbf{Pareto optimization:} 
Due to the difficulty in finding a global optimal solution for multiple directions, a common approach is to decompose multi-constraint optimization problems into sub-problems and search and seek different gradient descent directions.
Inspired by~\cite{dyankov2019multi}, we optimize the denoising results through a set of Pareto optimal solutions $\langle\epsilon_{\theta}^{n},\epsilon_{\theta}^{c},\epsilon_{\theta}^{s}\rangle$, known as the Pareto front.
Based on the Kahn-Kuhn-Tucker (KKT) condition, we calculate the KL-divergence between the generated samples and the conditions to derive the gradient direction,
\begin{equation}
\bigtriangledown d=-[\sum_{i\in \mathcal{B}}\lambda_{i} \mathcal{D}_{KL}(\epsilon_{\theta}^{n},\epsilon^{n})+\sum_{j\in \mathcal{B}}\beta_{j} \mathcal{D}_{KL}(\epsilon_{\theta}^{c},\epsilon^{c})+\sum_{m\in \mathcal{B}}\omega_{m} \mathcal{D}_{KL}(\epsilon_{\theta}^{s},\epsilon^{s})],
\label{eq6}
\end{equation}
where $\forall \lambda_{i},\beta_{j},\omega_{m}\geq 0$ are the Lagrange multipliers for the linear inequality constraints, s.t. $\sum_{i}\lambda_{i}+\sum_{j}\beta_{j}+\sum_{m}\omega_{m}=1$.
In this way, we base the decision on the objective space and constraint space, and no longer on the parameter space.
The Pareto optimization procedure is shown in Alg.~\ref{alg1}.

\begin{algorithm}[htbp]
\caption{Pareto optimization}\label{alg1}
\begin{algorithmic}[1]
\REQUIRE A semantic set $z_{t}=[n_{t}^{(i)},c_{t}^{(j)},s_{t}^{(m)}]_{i,j,m}$ based on the training batch $\mathcal{B}$
\ENSURE The solution $\epsilon_{\theta}^{\prime}$ for all subproblems with different trade-offs $\{n_{t},c_{t},s_{t}\}$
\STATE Initialize model parameters $\epsilon_{\theta}^{n}$, $\epsilon_{\theta}^{c}$, $\epsilon_{\theta}^{s}$ for multi-branch denoisers
\FOR {$i,j,m$ in batch $\mathcal{B}$}
\STATE Find the optimal parameters $\theta^{(n)}_{\mathcal{B}}$, $\theta^{(c)}_{\mathcal{B}}$, $\theta^{(s)}_{\mathcal{B}}$ using gradient-based method
\STATE Obtain $\lambda_{i}\geq 0$, $\beta_{j}\geq 0$, $\omega_{m}\geq 0$ by solving
\STATE \hspace{2em} $\arg\min -\frac{1}{2}||\sum_{i}\lambda_{i} \mathcal{D}_{KL}(\epsilon_{\theta}^{n},\epsilon^{n})+\sum_{j}\beta_{j} \mathcal{D}_{KL}(\epsilon_{\theta}^{c},\epsilon^{c})+\sum_{m}\omega_{m} \mathcal{D}_{KL}(\epsilon_{\theta}^{s},\epsilon^{s})||^{2}$
\STATE update the parameters $\epsilon_{\theta}^{\prime}=\epsilon_{\theta}+\bigtriangledown d_{\mathcal{B}}$
\ENDFOR
\end{algorithmic}
\end{algorithm}

The branch denoisers $\epsilon_{\theta}^{n}$, $\epsilon_{\theta}^{c}$, $\epsilon_{\theta}^{s}$ can approximate the optimal solution independently since the feature is fused in the subsequent channel mix blocks.
According to the law of large numbers, ensuring a sufficient quantity of batches $\mathcal{B}$ is a prerequisite to approximate the global optimum effectively.

\section{Experiments}
In this section, we conduct music generation experiments to verify the effectiveness of our method and study its characteristics through ablations.
The case studies offer an extended analysis of the structural regularity of generated samples.
In the experiment, we retrained the comparisons and reported the average of 5 computed values.
The MIDI scores and retrievable semantic information are converted by transcription software.
The model configurations are shown in Appendix B.

\textbf{Datasets:} 
We first validated the FSL-V2 module on the GuitarPro dataset to ascertain its accuracy in semantic extraction.
Subsequently, we trained the generative model on the Bread-midi dataset and validated it using pitch-, rhythm- and structure-based metrics.
The dataset details are as follows:
(1) GuitarPro~\cite{wu2023power} dataset comprises 3,188 semantic sections, categorized into intro, verse, chorus, bridge, outro, and 5 other categories;
(2) Bread-midi~\cite{bread-midi} dataset contains 946,909 MIDI files and is by far the most comprehensive MIDI dataset on the web.

\subsection{Fragmentation evaluation}
The accuracy of semantic fragmentation is a prerequisite for conditional music generation, so we validated the modified FSL-V2 module on the GuitarPro dataset.
To eliminate configuration bias, we trained multiple sets of parameters by left-to-right (L2R) and global padding strategies.
In addition, we performed a comparison between MIDI and event-based notation, with and without SSIM index.
Predicted sections are measured by 4 common metrics, Intersection over Union (IoU), Unweighted Average Recall (UAR), Root Mean Squared Error (RMSE), and determination coefficient ($R^{2}$).

\begin{table}[!htbp]
\centering
\caption{Comparison of pre-trained fragmentation modules on the GuritarPro dataset.}
\label{table1}
\begin{subtable}{0.48\columnwidth}  
\resizebox{1\columnwidth}{!}{
\begin{tabular}{lccccc}
\hline
\textbf{FSL-v1}&Padding&IoU&UAR&RMSE&$R^{2}$\\
\hline
\multirow{2}{*}{\makecell[c]{\scriptsize{-MIDI without SSIM}}}&\scriptsize{L2R}&0.912&0.892&6.641&0.666\\
~&\scriptsize{Global}&0.912&0.922&6.080&0.683\\
\hline
\multirow{2}{*}{\makecell[c]{\scriptsize{-MIDI with SSIM}}}&\scriptsize{L2R}&0.931&0.915&5.856&0.633\\
~&\scriptsize{Global}&0.946&0.956&5.492&0.621\\
\hline
\end{tabular}} 
\end{subtable}  
\begin{subtable}{0.48\columnwidth}  
\resizebox{1\columnwidth}{!}{
\begin{tabular}{lccccc}
\hline
\textbf{FSL-v2}&Padding&IoU&UAR&RMSE&$R^{2}$\\
\hline
\multirow{2}{*}{\makecell[c]{\scriptsize{-REMI with SSIM}}}&\scriptsize{L2R}&0.944&0.935&4.273&0.861\\
~&\scriptsize{Global}&0.953&0.973&3.730&\textbf{0.951}\\
\hline
\multirow{2}{*}{\makecell[c]{\scriptsize{-CP words with SSIM}}}&\scriptsize{L2R}&0.967&0.924&2.813&0.927\\
~&\scriptsize{Global}& \textbf{0.971}&\textbf{0.986}&\textbf{2.394}&0.944\\
\hline    
\end{tabular}}
\end{subtable}
\end{table}

\begin{figure}[htbp]
\centering
\includegraphics[width=0.98\columnwidth]{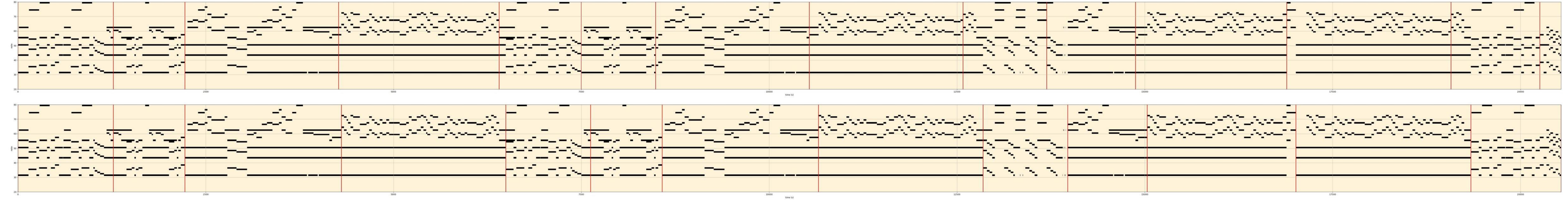}
\caption{Example of fragmentation difference between FSL-v1 (upper) and FSL-v2 (lower).}
\label{fig4}
\end{figure}

Tab.~\ref{table1} records the fragment results, with the best results in bold.
We can observe that, (1) usage of event-based notations can substantially improve IoU and RMSE; (2) global padding strategies can provide higher UAR compared to the L2R strategy; (3) usage of SSIM ensures a higher determination coefficient.
The FSL-v2 module outperforms the previous version on almost all metrics, thus providing a guarantee of correct semantics for subsequent experiments.
We can easily recognize the differences in visualized notes before and after modifying the fragmentation module, as shown in Fig.~\ref{fig4}, with the predicted sections highlighted in red.

\subsection{Generation evaluation}
We compare the language- and diffusion-based models with our Music-Diff architecture through the music generation task.
Considering language-based models, we retrained PopMusicTransformer~\cite{HuangY20}, Transfromer-GANs~\cite{Transformer-GANs}, and Jazz Transformer~\cite{Jazz-Transformer} as comparisons.
Likewise, we retrained diffusion-based models, such as SCHmUBERT~\cite{PlasserPW23}, Diffusion-LM~\cite{LiTGLH22}, and DiffuSeq~\cite{GongLF0K23}.
The results are recorded in Tab.~\ref{table2}, with detailed metrics provided in Appendix C.

\renewcommand\arraystretch{1.2}
\begin{table*}[!htbp]
\centering
\caption{Comparison of generation performance.
Upward or downward trends relative to the data set are denoted by \textcolor{red}{$\uparrow$} and \textcolor{teal}{$\downarrow$}, respectively, and trends of more than 10\% are denoted by \textcolor{red}{$\uparrow\uparrow$} and \textcolor{teal}{$\downarrow\downarrow$}.}
\label{table2}
\resizebox{0.98\columnwidth}{!}{
\begin{tabular}{c|c|cccc|cccc|ccc}
\hline
\multirow{2}{*}{\textbf{Samples}}&\multirow{2}{*}{\textbf{PPL}}&\multicolumn{4}{c|}{\textbf{\makecell[c]{Pitch-based metrics}}}&\multicolumn{4}{c|}{\textbf{\makecell[c]{Rhythm-based metrics}}}&\multicolumn{3}{c}{\textbf{\makecell[c]{Structure-based metrics}}}\\
\cline{3-13}
~&~&\textbf{PCU}&\textbf{TUP}&\textbf{PR}&\textbf{APS}&\textbf{ISR}&\textbf{PRS}&\textbf{IOI}&\textbf{GS}&\textbf{PCH}&\textbf{CPI}&\textbf{SI}\\
\hline
Training Set (overall)&-&7.066&56.67&60.36&11.61&0.819&0.427&0.079&0.908&2.911&0.964&0.163\\
\hline
MusicTransformer&1.201&6.140\textcolor{teal}{$\downarrow\downarrow$}&54.58\textcolor{teal}{$\downarrow$}&61.71\textcolor{red}{$\uparrow$}&11.62\textcolor{red}{$\uparrow$}&0.831\textcolor{red}{$\uparrow$}&0.493\textcolor{red}{$\uparrow\uparrow$}&0.071\textcolor{teal}{$\downarrow$}&0.918\textcolor{red}{$\uparrow$}&2.539\textcolor{teal}{$\downarrow\downarrow$}&0.990\textcolor{red}{$\uparrow$}&0.091\textcolor{teal}{$\downarrow\downarrow$}\\
Transformer-GANs&1.177&6.124\textcolor{teal}{$\downarrow\downarrow$}&55.10\textcolor{teal}{$\downarrow$}&62.75\textcolor{red}{$\uparrow$}&11.90\textcolor{red}{$\uparrow$}&0.814\textcolor{teal}{$\downarrow$}&0.304\textcolor{teal}{$\downarrow\downarrow$}&0.090\textcolor{red}{$\uparrow\uparrow$}&0.919\textcolor{red}{$\uparrow$}&3.461\textcolor{red}{$\uparrow\uparrow$}&0.997\textcolor{red}{$\uparrow$}&0.123\textcolor{teal}{$\downarrow\downarrow$}\\
Jazz Transformer&1.122&6.783\textcolor{teal}{$\downarrow$}&61.33\textcolor{red}{$\uparrow$}&64.26\textcolor{red}{$\uparrow$}&11.94\textcolor{red}{$\uparrow$}&0.818\textcolor{teal}{$\downarrow$}&0.322\textcolor{teal}{$\downarrow\downarrow$}&0.119\textcolor{red}{$\uparrow\uparrow$}&0.898\textcolor{teal}{$\downarrow$}&2.294\textcolor{teal}{$\downarrow\downarrow$}&0.386\textcolor{teal}{$\downarrow\downarrow$}&0.087\textcolor{teal}{$\downarrow\downarrow$}\\
\hline
SCHmUBERT(D3PM)&1.345&5.329\textcolor{teal}{$\downarrow\downarrow$}&25.58\textcolor{teal}{$\downarrow\downarrow$}&34.08\textcolor{teal}{$\downarrow\downarrow$}&8.895\textcolor{teal}{$\downarrow\downarrow$}&0.761\textcolor{teal}{$\downarrow$}&0.398\textcolor{teal}{$\downarrow$}&0.198\textcolor{red}{$\uparrow\uparrow$}&0.932\textcolor{red}{$\uparrow$}&2.237\textcolor{teal}{$\downarrow\downarrow$}&0.949\textcolor{teal}{$\downarrow$}&0.037\textcolor{teal}{$\downarrow\downarrow$}\\
Diffusion-LM&1.261&6.252\textcolor{teal}{$\downarrow\downarrow$}&45.68\textcolor{teal}{$\downarrow\downarrow$}&52.58\textcolor{teal}{$\downarrow\downarrow$}&8.880\textcolor{teal}{$\downarrow\downarrow$}&0.798\textcolor{teal}{$\downarrow$}&0.420\textcolor{teal}{$\downarrow$}&0.143\textcolor{red}{$\uparrow\uparrow$}&0.913\textcolor{red}{$\uparrow$}&2.334\textcolor{teal}{$\downarrow\downarrow$}&0.973\textcolor{red}{$\uparrow$}&0.091\textcolor{teal}{$\downarrow\downarrow$}\\
DiffuSeq&1.254&7.548\textcolor{red}{$\uparrow$}&44.32\textcolor{teal}{$\downarrow\downarrow$}&53.48\textcolor{teal}{$\downarrow\downarrow$}&9.765\textcolor{teal}{$\downarrow\downarrow$}&0.812\textcolor{teal}{$\downarrow$}&0.438\textcolor{red}{$\uparrow$}&0.137\textcolor{red}{$\uparrow\uparrow$}&0.924\textcolor{red}{$\uparrow$}&2.855\textcolor{teal}{$\downarrow$}&0.974\textcolor{red}{$\uparrow$}&0.095\textcolor{teal}{$\downarrow\downarrow$}\\
\hline
Music-Diff&1.113&9.861\textcolor{red}{$\uparrow\uparrow$}&71.24\textcolor{red}{$\uparrow\uparrow$}&80.82\textcolor{red}{$\uparrow\uparrow$}&8.431\textcolor{teal}{$\downarrow\downarrow$}&0.871\textcolor{red}{$\uparrow$}&0.477\textcolor{red}{$\uparrow\uparrow$}&0.078\textcolor{teal}{$\downarrow$}&0.908-&3.088\textcolor{red}{$\uparrow$}&0.999\textcolor{red}{$\uparrow$}&0.189\textcolor{red}{$\uparrow\uparrow$}\\
\hline
\end{tabular}}
\end{table*}

\textbf{Language vs. diffusion models:}
For the pitch metrics, the language model is closer to the original dataset, whereas the diffusion model shows a substantial decrease in all aspects, suggesting a lower pitch diversity and a narrowing of the span between pre- and post-pitch.
For the rhythmic metrics, although there are more non-zero entries in the diffusion models, there is a significant improvement in the consistency and clarity of rhythm in the diffusion model, which is reflected in IOI and GS.
Compared to diffusion models, language models excel in maintaining structural consistency, but both models perform poorly in long-term structural metrics such as SI.
Above, we conclude that,
(1) For language models, while the diversity metrics also reveal shortcomings, the more significant deficiency lies in maintaining consistency of long-term structures, which is attributed to the inherent maximum length limitation and the accumulation of errors;
(2) Although diffusion models can directly learn sample distributions, the increased diversity in high-dimensional space enhanced by existing models through discrete-continuous mapping is not reflected in the generated samples.

\textbf{Our performance:}
Our architecture has excellent PCU (9.86), TUP (71.24), PR (80.82), PCH (3.088), and SI (0.189), as well as other slightly improved performance.
The advantages of our model are: 
(1) Greater pitch diversity, which may be attributed to the joint semantic pre-training method providing rich semantic-based tokens, or perhaps the noise perturbation enables the model to explore more possibilities in the high-dimensional semantic space;
(2) Excellent rhythmic and structural coherence, perhaps benefits to semantic information integration from the multi-branch denoiser, and the deterministic constraints provided by the prompt-learning for compositional rules.

\subsection{Ablations}
\begin{figure}[htbp]
\centering
\includegraphics[width=0.98\columnwidth]{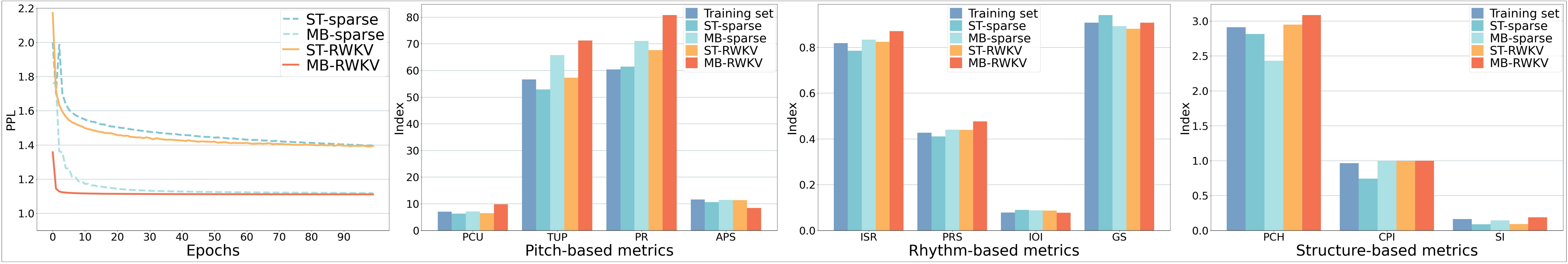}
\caption{Comparison of samples generated using different backbones.}
\label{fig5}
\end{figure}

\begin{figure}[htbp]
\centering
\includegraphics[width=0.98\columnwidth]{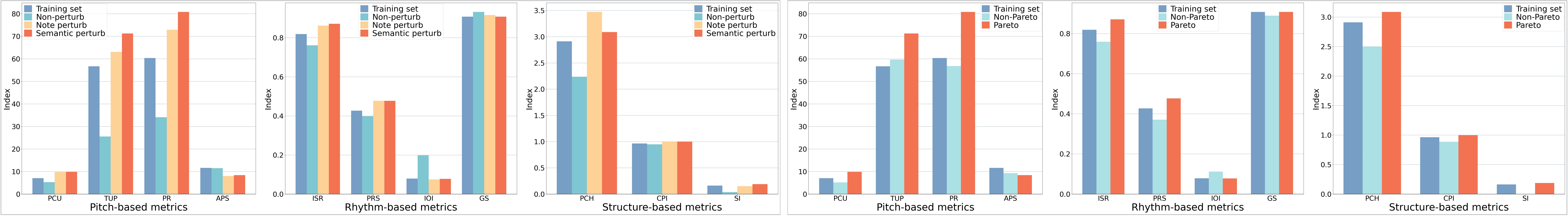}
\caption{Ablations of perturbations configuration (left) and Pareto optimization (right).}
\label{fig6}
\end{figure}

We analyzed the performance of our model with different configurations, aiming to test three aspects:
(1) optimal configuration of backbone;
(2) optimal choice with perturbations;
(3) validity of Pareto optimization.
The results are demonstrated in Fig.~\ref{fig5}, and Fig.~\ref{fig6}.

We observed that the Multi-Branch (MB) architecture significantly reduces the PPL compared to Single-Thread (ST).
Furthermore, MB architectures achieve better results across almost all metrics than ST architectures, especially PCU, TUP, PR (pitch diversity); ISR, PRS (rhythm consistency); and SI (long-term musical normality).
The performance of the RWKV backbones is slightly superior to sparse attention, with its notable contribution being the improvement in convergence speed.

Further analysis in Fig.~\ref{fig6}(left) shows that noise perturbation significantly improves sample diversity, more so at the semantic level than at the note level, as evidenced by the significant differences in the TUP, PR, ISR, and PCH metrics. 
Through the ablation study depicted in Fig.~\ref{fig6}(right), it is evident that Pareto optimization markedly enhances almost all pitch-based indicators, resulting in generated samples that exhibit greater rhythm and structural consistency with the training set.
These observations validate the practicality of Pareto optimization in controlling music style.

\subsection{Case studies}
\begin{figure}[htbp]
\centering
\includegraphics[width=0.98\columnwidth]{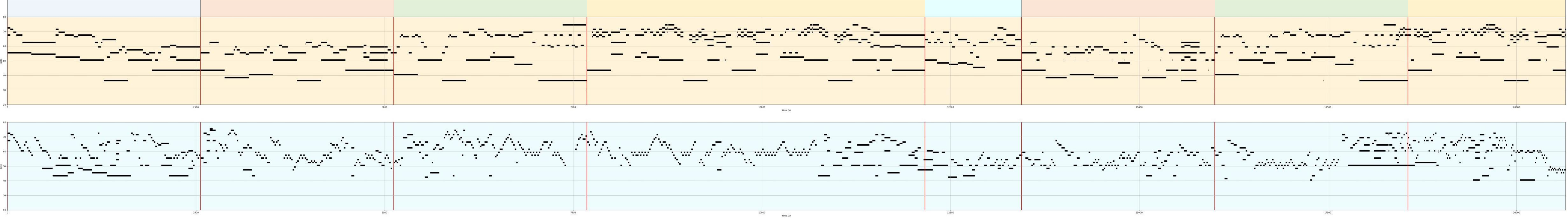}
\caption{Example of a visualized MIDI score, with blue (bottom) representing a sample generated by semantic prompts from the orange (top) reference sample.}
\label{fig7}
\end{figure}

We also assess the aesthetic appeal of the musical instances by visualizing the generated music. 
Fig.~\ref{fig7} demonstrates a sample generated using semantic prompts, with different colored boxes to distinguish the prompt semantics.
We observe that the melodies generated by our model, although different from the reference sample, possess a strong overall regularity, especially in the rise and fall of consecutive notes, which dispels our concerns about compositional normality.

\begin{figure}[htbp]
\centering
\includegraphics[width=0.98\columnwidth]{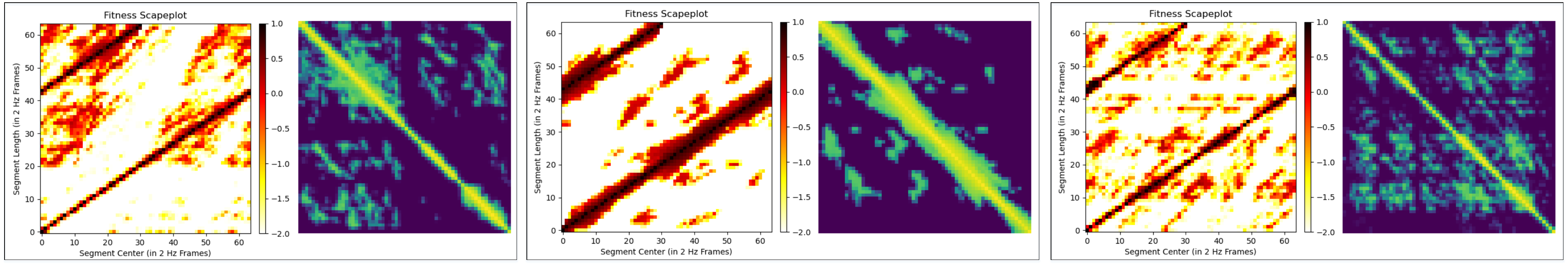}
\caption{FS and SSM plots for samples generated by different models.}
\label{fig8}
\end{figure}

\textbf{Structural analysis:} 
The difficulty in generating music that sounds different from random notes lies in reproducing the complex structure of the music.
Therefore, we analyzed the sample structure according to Fitness Scape (FS) and Self-Similarity Matrices (SSM), primarily aiming to investigate contextual correlations and identify recurring structures.
Fig.~\ref{fig8} illustrates the analysis of the language-based (MusicTransformer), DDPMs-based (SCHmUBERT), and our diffusion model.

The FS plot shows that our model regularly generates melodies with high structural similarity.
As a comparison, the language model loses structural relevance when generated notes exceed the preset maximum length.
With SSM plots, our model exhibits periodic structural properties that are overwhelmingly superior in long-term correlations.
The advantages of our rule-based melodies are also reflected in subjective auditory tests, making the generated music more realistic.

\section{Conclusions}
Most researchers focus on verbalizing music while ignoring that musical notes themselves are continuous in the frequency domain.
As a unified architecture for symbolic music generation, we introduced the JSP approach to perform multivariate perturbations and the Symb-RWKV model to recover joint distributed noise.
The resulting architecture is trained on musical structures to enhance the composition regularity, thereby reducing the risk of semantic information loss.
We conducted a series of music generation experiments to better understand the advantages of diffusion-based models and the optimal parameters of denoiser.
Notably, we found that 
(1) Multivariate perturbation of musical notes has significant advantages in increasing the diversity of generated music, including our advocacy for modeling "never-used" notes;
(2) Compared to multivariate noise prediction, the score-matching method can only express the difference in the statistical distribution of the note itself, but it has a poor ability to express the higher-order semantics, which is reflected in maintaining the note-level diversity but lacks structural consistency of the generated samples.

\textbf{Future works:}
We plan to generalize our approach to multiple instruments collaborating for polyphonic music generation.
To achieve a symphonic ensemble, our next step involves dividing and discerning the tonal and compositional characteristics of individual instruments. 

\bibliographystyle{unsrtnat}
\bibliography{neurips_2024.bib}


\appendix
\label{sec:appendix}
\newpage

\section{Symbol Explanation}
The mathematical symbols defined in this paper are listed in the Tab.~\ref{table3}:

\renewcommand\arraystretch{1.2}
\begin{table}[htbp]
\centering
\caption{Summary of the Notations}
\label{table3}
\resizebox{0.98\columnwidth}{!}{
\begin{tabular}{p{0.15\textwidth}<{\centering}p{0.85\textwidth}}
\hline
\multicolumn{1}{c}{\textbf{Symbols}}&\textbf{Meaning}\\
\hline
$x_{(\cdot)}$&A music symbols sequence expressed in MIDI or event-based notation.\\
$z_{n}$,$z_{c}$,$z_{s}$&The embedded representations are fragmented into note, chord, and section labels.\\
$i$,$j$,$m$&The number of notes, chords, and sections in the symbol sequence.\\
$Cand_{m}$&The $m$-th candidate windows for fragmentation operation.\\
$C$,$\eta$&The center position and size of the padded candidate windows.\\
$\mu$,$\sigma$&The mean and standard deviations of pitches within candidates.\\
$\delta$&The added constant to avoid instability when the denominator is zero.\\
$p_{cls}$&The classification probabilities of candidate windows.\\
$\mathcal{L}_{cls}$,$\mathcal{L}_{reg}$&The classification, regression loss terms in the fragmentation module.\\
$\varepsilon$&The encoders for embedding the note, chord, section representations.\\
$t$,$T$&The timestamps $t$ of the noise addition steps $T$.\\
$\epsilon$&Fitted noise observations.\\
$\gamma(\cdot)$,$g(\cdot)$&The drift and diffusion coefficients obeying a Gaussian distribution.\\
$\alpha$&Constants $\alpha$ are hyperparameters.\\
$\theta$&The parameter of denoising model.\\
$R$,$W$,$K$,$V$&Receptance, weighted, key, value vectors transformed from parameter matrices.\\
$\varphi$,$\phi$&Serial number and the total number of heads of the attention mechanism.\\
$\xi$&The sigmoid operation.\\
$o_{t}$&The output of the temporal or channel mix blocks.\\
$\lambda$,$\beta$,$\omega$&The weights of the Pareto optimization.\\
$\mathcal{B}$&The mini-batch in the training.\\
\hline
\end{tabular}}
\end{table}

\section{Details of model training}
We pre-trained a Joint Semantic Pre-training (JSP) module for constructing musical progression relations of note and semantics and trained different branches of the denoiser through the RWKV attention mechanism.

\textbf{Joint semantic pre-training:}
The JSP module was trained from scratch rather than initialized with migrated weights.
JSP module is built on a unified embedding layer, where the vocabulary includes note pitches and section semantics.
Due to different contextual environments, their maximum lengths also vary. 
Based on experience, the context sequences we indexed consist of 2048 notes, 256 chords, and 32 sections.
Since the semantics are sparse vectors relative to the notes, we uniformly sample the semantic distribution from the notes via the transfer functions $p(c,n)$ and $p(s,c)$.
The two-by-two contrastive modules, JSP-CN and JSP-SC, are constructed to predict paired relationships between 128 pitches, 48 chords, and 10 sections.
The model solely employs linear projection to map each semantic into the embedding space, thereby eliminating nonlinear projections in the contrastive embedding space.
Since there is no multimodal problem, the training details in JSP are simplified compared to CLIP~\cite{RadfordKHRGASAM21}.

\textbf{RWKV backbone:}
Our noise prediction backbones use parameters of 1.60 $B$, which is calculated as $13D^{2}L+468DL+4D+2DV$, where number of layers $L$=24, attention head $H$=32, and hidden dimension $D$=2048.
Based on the reference length, we set the maximum number of response tokens to 2048.
The batch size is set to 64 and the epoch number uses a non-default value to allow for early stopping.
We adopt the strategy of declined learning rate, i.e., the initial learning rate is $4e-4$ and will decrease to 60\% of the former every 5 epochs, and stop the training process if the loss reduction is less than $5*10^{-5}$ in 10 consecutive epochs.
In the optimizer, we use default parameters, including the $eps = 1e-8$ term in the denominator to enhance numerical stability during gradient computation, and $betas = (0.9, 0.99)$ coefficients for the running averages of gradient squares. 
As we utilize the largest existing MIDI dataset, the solution space becomes significantly large, and instances of gradient explosion are nearly non-existent. 
Therefore, we abandon the use of gradient clipping, i.e. $grad\_norm\_clip = 1.0$.
Since we learned to generate output by prompting, we eliminated the warm-up token and final token settings.
The temperature parameter that controls the range of the logits in the Gumbel Softmax is directly optimized during training as a log-parameterized multiplicative scalar to avoid turning into a hyper-parameter.

\section{Computation cost}
Since the complexity of deep learning frameworks depends on the used baselines, we compare the computation cost as follows, 
(1) the language models, e.g. MusicTransformer and JazzTransformer;
(2) the diffusion models, e.g. DiffuSeq;
(3) our used backbone RWKV.
Tab.~\ref{table4} shows the computation cost of comparing the models on two training datasets, using the metrics of parameter size and cost per batch time.

\renewcommand\arraystretch{1.2}
\begin{table}[htbp]
\scriptsize
\centering
\caption{Comparison of the computational complexity.}
\label{table4}
\resizebox{0.6\columnwidth}{!}{
\begin{tabular}{ccccc}
\hline
\multirow{3}{*}{\textbf{Baselines}}&\multicolumn{2}{c}{\textbf{\makecell[c]{GuitarPro}}}&\multicolumn{2}{c}{\textbf{\makecell[c]{Bread-midi}}}\\
\cmidrule(r){2-3}\cmidrule(r){4-5}
\textbf{~}&\textbf{\makecell[c]{Para size\\(MB)}}&\textbf{\makecell[c]{Cost time\\(s / batch)}}&\textbf{\makecell[c]{Para size\\(MB)}}&\textbf{\makecell[c]{Cost time\\(s / batch)}}\\
\hline
MusicTransformer& 273.6 &4.51&296&4.60\\
JazzTransformer & 471&4.14&482&4.43\\
DiffuSeq &1467&32.84&1539&33.61\\ 
RWKV-music&78.31&2.17&78.42&2.98\\ 
Music-Diff &103.4&3.21&101.94&3.58\\
\hline
\end{tabular}}
\end{table}

Obviously, the DiffuSeq models are more complex than other models.
However, using the same configuration, RWKV backbone takes less memory and has a faster runtime than the canonical Transformer model.
Given the sample length of $L$, the computational complexity of the post-optimized Transformer is $O\left(L \log L\right)$.
The time complexity of RWKV is close to linear, i.e. $O(L)$
Due to the added parallel branches and shared mix blocks, our Music-Diff increase the computational complexity to approximately $O\left((1/3+2\mathcal{N}_{head}/3)*L\right)$

\textbf{Platform:} 
All experiments were trained and tested on an Nvidia A100 80 GB GPU.

\section{Evaluation metrics}
We examine multiple aspects of the experimental sample through four sets of quantitative metrics~\cite{wu2023power}:

(1) Probability-based metrics:
\textbf{PPL} (Perplexity) is a common metric for evaluating the distribution of generated samples, reflecting the degree of match to the training set.

(2) Pitch-based metrics:
\textbf{PCU} (Pitch category unique) and \textbf{TUP} (Total number of unique pitches) indicate the number of pitch classes and the total number of pitches per bar;
\textbf{PR} (Pitch range) represents the average difference between the highest and lowest pitches in semitones;
\textbf{APS} (Average pitch by semitones) represents the average semitone interval between two consecutive pitches.
These metrics give a sense of the pitch diversity (PCU and TUP) and pitch variation (PR and APS) of the music samples.

(3) Rhythm-based metrics:
\textbf{ISR} (ratio of nonzero entries) indicates the rhythmic tightness of the sample on the global time scale;
\textbf{PRS} (ratio of pitches with time steps greater than 4 to the total number of pitches in a sample) reflects the percentage of notes of different lengths, such as dichotomous notes and quarter notes, in the sample.
\textbf{IOI} (Interonset-interval) represents the time between two consecutive notes and reflects the rhythmic similarity between the generated music and the original sample in a localized range.
\textbf{GS} (Grooving pattern similarity) indicates the similarity of grooving patterns between adjacent bars, with higher values indicating that a sample possesses a clear rhythm.
Rhythm-based metrics focus on the consistency of note length on the time step.

(4) Structure-based metrics:
\textbf{PCH} (Pitch-class histogram) represents an octave-independent representation of the 12-dimensional pitch content, and measures pitch stability over shorter time scales;
\textbf{CPI} (Chord progression irregularity) measures harmony consistency within a sample;
\textbf{SI} (Structures indicators) detects the presence of repeated structures within a specific time range.
These metrics describe structural events, and a large value of these metrics indicates the presence of multiplexed note patterns, chords, or melodies within the detection range.
\end{document}